 \documentclass[12pt]{iopart}

\usepackage{iopams,bm,graphics,graphicx,epsfig,color}

\begin{document}

\title[Generic bounds on Dipolar gravitational radiation]{Generic bounds on dipolar gravitational radiation from inspiralling compact binaries}

\author{K. G. Arun$^{1}$ }

\ead{kgarun@cmi.ac.in}
\address{
$^1$ Chennai Mathematical Institute, Siruseri, 603103, India.\\
}

\begin{abstract}
Various alternative theories of gravity predict dipolar gravitational radiation in addition to  quadrupolar radiation.  We show that gravitational wave (GW) observations of inspiralling compact binaries can put interesting constraints on the strengths of the dipole modes of GW polarizations. We put forward a physically motivated gravitational waveform for dipole modes,  in the Fourier domain, in terms of two parameters: one which captures the relative amplitude of the dipole mode with respect to the quadrupole mode ($\alpha$) and the other a dipole  term in the phase ($\beta$). We then use this two parameter representation to discuss typical bounds on their values using  GW measurements. We obtain the expected bounds on the amplitude parameter $\alpha$ and the phase parameter $\beta$ for Advanced LIGO (AdvLIGO) and Einstein Telescope (ET) noise power spectral densities using Fisher information matrix. AdvLIGO and ET may at best bound $\alpha$ to an accuracy of $\sim10^{-2}$ and $\sim10^{-3}$ and $\beta$ to an accuracy of $\sim10^{-5}$ and $\sim10^{-6}$ respectively.
\end{abstract}

\maketitle

\section{Introduction and summary}
General Relativity (GR) is the most empirically successful theory of relativistic gravity to date. Despite its success, there is strong motivation to test it in regimes where it has not been tested. Every relativistic theory of gravity has its own predictions about the nature and properties of gravitational waves (GWs). Hence observations of GWs from various astrophysical phenomena can be used to test various theories of gravity including GR~(see e.g. \cite{Wthexp,Will05LivRev,SathyaSchutzLivRev09}). Among the various sources of GWs, inspiralling compact binaries are the most promising, both for the first detection as well
as for testing alternative theories of gravity. This is because, for these sources
one can very accurately compute the gravitational waveforms within GR.


From a GW point of view, the modifications of GR can happen in two ways: one is in the {\it generation} of the waves, because of the presence of other fields like scalars or vectors which couple to the metric; and the second is in the effects of propagation of the waves. The major difficulty in analysing the effects of alternative theories of gravity in the context of GWs is that unlike GR, where very precise waveforms are available for inspiralling compact binaries~(see e.g.\cite{BDEI04, Bliving}), theoretical progress in modelling the GW emission within the framework of alternative theories of gravity is scanty, apart from scalar-tensor theories (see e.g.~\cite{CSreview09} for a review of Chern-Simons theory
and \cite{YPC11,HealyST11} for some recent works on scalar tensor theories). Hence one has to come up with parametrized descriptions of the waveform within different alternative theories of gravity.
For example, Will discussed model gravitational waveforms in the case of 
scalar tensor theories~\cite{WillBD94} (see also \cite{KKS95}) and massive graviton theories~\cite{Will98}. The waveforms could be expressed in terms of a one-parameter deviation from GR in the GW phasing in both cases. He also discussed the expected bounds on these parameters from various GW observations. These estimates were revised later accounting for various physical effects such as spin and using more up-to-date waveforms and detector noise characteristics~\cite{KKS95,WillYunes04,BBW05a,AW09,SW09,YagiTanaka09a,YagiTanaka09b,KeppelAjith10,MYW11}. 

There have also been attempts to write down generic gravitational waveforms in terms of the differences in the phasing coefficients, and to study the ability of GW detectors to measure a few of them. This amounts to performing consistency tests of post-Newtonian (PN) theory within GR in the mass plane of the binary components~\cite{AIQS06a,AIQS06b,MAIS10}. More recently, Yunes and Pretorius proposed a general parametrization of gravitational waveforms called the Parameterized-Post-Einsteinian (ppE) formalism~\cite{YunesPretorius09,PPE2011}. 

A generic metric theory of gravity has six modes of GW polarizations as opposed to the two predicted by GR~\cite{EardleyEtal73a,EardleyEtal73b}. If these additional modes were to be detected by GW experiments, it would rule out GR.  Alternatively, GW observations
can put stringent constraints on the parameters governing these additional polarization modes. If no assumptions are made about the source geometry or the emitted waveform, one requires six independent GW data streams plus knowledge of the source direction to measure or constrain all six modes of GW polarizations.  This would mean a network of six independent ground based-GW detectors which does not seem plausible in the immediate future. 

In this paper we show that one can constrain a {\it restricted class} of modes of polarization, which correspond to dipolar GW emission, using  interferometric measurement of the GW signal. {We discuss the bounds on a class of theories which predict dipolar GW radiation in addition to the quadrupolar GW emission  similar to GR. 
We propose a physically motivated waveform model of dipolar GW modes of polarization, containing two free parameters, that could arise in a theory of gravity with a vector field or scalar field in addition to the usual metric. We then use this two-parameter description (in addition to the standard GR parameters) to discuss the bounds possible from ground-based GW interferometers. {In addition to scalar-tensor theories like Brans-Dicke theory which predicts dipolar GW emission, the recently discussed theories with vector fields like the TeVeS theory~\cite{Bekenstein04,Sagi10} and Einstein-Aether theory~\cite{Jacobson08}, and the Modified Gravity theory (MOG)~\cite{Moffat05}  also predict dipolar GWs.}

The paper is organized in the following way. Sec.~\ref{sec:assumptions} lists the assumptions we are making to obtain a gravitational waveform, including the contributions from the dipolar modes.  The Fourier-domain waveform is derived in Sec.~\ref{sec:derive}. The details of the Fisher matrix analysis are presented in Sec.~\ref{sec:Fisher}, and the results are discussed in Sec.~\ref{sec:results}. Conclusions are provided in Sec.~\ref{sec:conclusion}.
\section{Assumptions}\label{sec:assumptions}
In order to write down a parametrized waveform which captures both the quadrupolar and dipolar GWs, we make the following set of assumptions.
\begin{enumerate}
\item For the quadrupole part we use the GR waveform. In other words we completely ignore the corrections to the quadrupolar component of the waveform from the underlying theory of gravity. This is consistent with the theme of the paper where we focus on the dipole component of the gravitational waveform. One can include the corrections to the GR waveform following the general parametrizations given in \cite{AIQS06a,YunesPretorius09} which is outside the scope of this paper. 
\item The phasing of the dipolar GW mode will be proportional to the orbital phase instead of twice the orbital phase as for the quadrupolar tensor mode.  Further, because of a dipole component in the energy flux, the orbital phase evolution will also have dipole term similar to the
case of scalar-tensor theories~\cite{WillBD94}.  However, we will ignore higher harmonics of the orbital phase above the quadrupolar one.

\item 
From dimensional arguments, the frequency dependence of the phasing term from the dipole contributions should be $O(v^{-2})$ relative to the GR quadrupole contribution, and that of the amplitude should be $O(v^{-1})$ times the leading quadrupole term, where $v$ represents the characteristic source velocity. This is easy to understand intuitively, as in the total GW flux for the quadrupole and dipole terms differ by $({d}/{dt})^2\sim {\cal O} (v^2)$ whereas for the waveform they differ by ${d}/{dt}\sim {\cal O}(v)$. (For the multipolar structure of energy flux and waveform see  Eq. (12.2) of Ref.~\cite{BIJ02}  and (3.11) of \cite{ABIQ04} respectively. Though these papers discuss multipoles higher than the quadrupole, it is easy to infer the $v$ dependence of the dipole term from them.)
\item In the underlying theory which predicts dipolar radiation, there could also be monopole radiation.  However, by analogy with scalar-tensor theories, it is likely that such monopole contributions will begin at post-Newtonian order, since the monopole (the mass) is conserved at Newtonian order, and will thereby have the same scaling with $v$ as the quadrupole contributions. We have also not considered possible deviation of effective stress energy tensor of GWs from that of GR as discussed in Ref.~\cite{SteinYunes11}. 
\item We have assumed that the dipolar modes propagate at the velocity of light.
As mentioned earlier, introducing an additional massive graviton parameter~\cite{Will98}, one can probe  the non-null propagation of these modes. But we avoid this added complication in the present work.
\item In the waveform that we write down below, we introduce two parameters $\alpha$ and $\beta$, which capture the relative amplitude of the dipolar mode with respect to the GR mode and  an additional dipole term in the phasing formula, respectively. It is likely that either or both of these parameters will depend on the masses and
on the parameters determining the structure of the compact objects, as in
Brans-Dicke (BD) theory (see Eq.\ (90) of \cite{Foster08} for the case of Einstein-Aether theory). However, we will not attempt to include such dependences in the parameters $\alpha$ and $\beta$. But it should be noted that for objects like neutron stars,
these parameters may depend upon the equation of state.
\item Lastly, we have considered here single interferometric GW observations. But it can be extended to the case of GW detector network. 
\end{enumerate}
\section{Derivation of the phasing formula for dipolar modes}\label{sec:derive}
We start with a schematic expression for the detected strain at the detector output after projecting onto the response of the detector combining the GR and the dipole modes, writing them as
\begin{eqnarray}
h(t)&=& h_{\rm GR}(t)+ h_{\rm Dip}(t),\\
    &=& A_{\rm GR}(t) e^{2i\Psi(t)} + A_{\rm Dip}(t) e^{i\Psi(t)},\\
    &=& A\, v^2(t) \,e^{2i\Psi(t)}+ A\, \alpha \,v(t)\,e^{i\Psi(t)},
\end{eqnarray}
where $\Psi(t)$ is the time domain GW phase, $A_{\rm GR}(t)$ and $A_{\rm Dip}(t)$
are the time domain amplitudes of the GR and dipole modes, $\alpha$ is the parameter which captures the relative strength of the dipole mode with respect to the GR mode
and $v(t)$ is the characteristic instantaneous velocity of the binary.
Since the analysis is for a single detector, we have averaged over the antenna pattern functions.  This averaging yields numerical factors which have been
absorbed in the definition of the two amplitudes.

Using the stationary phase approximation (SPA) (see \cite{DIS00,YABW09} for detailed discussions and examples) one can obtain the Fourier-domain gravitational waveform. We follow the procedure developed by Van Den Broeck and Sengupta\cite{ChrisAnand06} to address the complication in computing the Fourier transform when higher multipolar components of the binary are to be taken into account.  As a simple first step, we include only the dipole and quadrupole harmonics, and the SPA leads to the expression
\begin{eqnarray}\label{eq:FDWF}
\tilde{h}(f)&=& {\tilde A}\frac{v_2^2(f)}{\sqrt {2{\dot F}(v_2)}}\,e^{2i\,{\tilde \Psi}(v_2)}+{\alpha\,{\tilde A}}\frac{v_1(f)}{\sqrt{{\dot F}(v_1)}}e^{i{\tilde \Psi}(v_1)}  \,,
\end{eqnarray}
where $v_k={(2\pi\,m\,f/k)}^{1/3}$, and $\dot{F}$ is the frequency sweep, which can be  written in terms of the
GR contribution $\dot{F}_{\rm GR}$ plus a  dipolar correction. Introducing the parameter $\beta$ for the dipolar correction, we express $\dot{F}$ in the form
\begin{eqnarray}
\dot{F}(v_k)&=&\dot{F}_{\rm GR}^{\rm Newt}\,\left[1+\beta\, v_k^{-2}+ {\cal O}(v_k^2) \right] \,,
\label{Fdot}
\end{eqnarray}
where $k=2$ for the quadrupole GR mode and $k=1$ for the dipolar  modes. Since we use the restricted waveform with Newtonian amplitude only the leading term in the ${\dot F}_{\rm GR}$ expression is required for the calculation, which we denote by $\dot{F}_{\rm GR}^{\rm Newt}$.
Substituting Eq.\ (\ref{Fdot}) into Eq.\ (\ref{eq:FDWF}) and keeping terms in the amplitude to linear order in $\alpha$ and $\beta$, we obtain
\begin{eqnarray}
\tilde{h}(f)
&\simeq& {\tilde A}\left[\frac{v_2^2}{\sqrt{2{\dot F_{\rm GR}^{\rm Newt}}(v_2)}}\left(1-\frac{\beta}{2}v_2^{-2}\right)\,e^{2i\,{\tilde \Psi}(v_2)}+\frac{\alpha\,v_1}{\sqrt{{\dot F_{\rm GR}^{\rm Newt}(v_1)}}}\,e^{i\,{\tilde \Psi}(v_1)}\right] \,.
\label{eq:VT}
\end{eqnarray}
In the above expression, the Fourier Domain phasing is given by
\begin{equation}
{\tilde \Psi}(v_k)={\tilde \Psi}^{\rm Newt}_{\rm GR}(v_k)\left(1+\overline{\beta}v_k^{-2}+\cdots\right),
\end{equation} 
where it is straightforward to show that $\overline{\beta}={-4\beta}/{7}$. The above expression is normalized to the Newtonian contribution to the phasing formula, $\Psi_{\rm GR}^{\rm Newt}$ and we account for the full GR phasing formula up to 3.5PN~\cite{DIS01,DIS02} in our calculations. 

 There is an important structural difference between the two terms which represent amplitude deviations in the final waveform. The term proportional to  $\alpha$ comes purely from the time-domain dipole mode polarization and its relative strength with respect to the GR polarization modes. The second term proportional to $\beta$ arises from the calculation of the Fourier domain waveform using the SPA (via ${\dot F}$) and this parameter is related to the strength of dipolar radiation component (energy flux) with respect to the quadrupolar radiation of GR. This parameter modifies the phasing of both the quadrupole and the dipole modes, as the effective energy balance will involve both the dipolar and the quadrupolar components of energy flux.

It is useful to see how this waveform maps onto the more generalized parametrizations mentioned earlier.
There are two leading contenders for general parametrizations of inspiral waveforms: the `parametrized tests' of PN theory (PTPN) by Arun et al~\cite{AIQS06a,AIQS06b,MAIS10} and the Parametrized post-Einsteinian (ppE) representation by Yunes and Pretorius~\cite{YunesPretorius09,PPE2011}. The proposal by Arun et al is to determine, from GW observations, various phasing coefficients and check for their consistencies in the mass plane. Their proposal, so far, has not been extended to include the deviations in amplitude w.r.t GR. Hence all that we can say is that there will be additional dipole contributions to the parametrized phasing expression in their formalism.

On the other hand the ppE accounts for typical amplitude and phase deviations from GR  in their parametrization.  For this comparison, we rewrite Eq.~(\ref{eq:VT}) as 
\begin{eqnarray}
{\tilde h}(f) &=& {\tilde{\cal A}_{\rm GR}}(f) \left(1-\frac{\beta}{2}\,v_2^{-2}\right) e^{2i\,{\tilde\Psi}^{\rm Newt}_{\rm GR}(v_2)\left[1+{\bar{\beta}}v_2^{-2}+\cdots\right]} + \delta {\tilde h}(v_1),\\
&=& {\tilde h}_{\rm GR} \left(1-\frac{\beta}{2}\,v_2^{-2}\right) e^{2i {\bar {\bar \beta}}v_2^{-2}} + \delta {\tilde h}(v_1),\label{eq:VT-re} 
\end{eqnarray}
where  ${\bar {\bar \beta}}={\tilde\Psi}_{\rm GR}^{\rm Newt}{\bar \beta}$ and $\delta{\tilde h(v_1)}$ is the amplitude term proportional to $\alpha$ in Eq.~(\ref{eq:VT}).
Comparing the above equation with Eq.~(1) of Ref.~\cite{YunesPretorius09} and ignoring the merger and ringdown part of the waveform, it is straightforward to see that the first term of Eq.~(\ref{eq:VT-re}) can be written down in a ppE form by choosing $\alpha_{\rm ppE}=-{\beta\over 2}\eta^{-2/5}$, $\beta_{\rm ppE}=2{\bar {\bar {\beta}}}\eta^{-2/5}$, $a=-2/3$ and $b=-2/3$. In the above expression, $\eta$ refers to the symmetric mass ratio which is the ratio  of reduced mass to the total mass.  However second term of Eq.~(\ref{eq:VT}) cannot be mapped to any corresponding term in ppE. This is because ppE, in its current form, does not incorporate harmonics other than the leading quadrupolar one. In other words,
the additional terms which arise here should be added to the ppE waveform when non-quadrupolar modes are included into their framework. 

\vskip 1cm
\section{Fisher matrix analysis}\label{sec:Fisher}
 Our goal now is to obtain the bounds on $\alpha$ and $\beta$ that would be possible with various GW detectors and for different types of sources. Using the waveform in Eq.~(\ref{eq:VT}), we perform a Fisher matrix analysis of the problem\footnote{See Ref.\cite{CF94} for one of the first applications of Fisher matrix in GW context and the basic theory involved.} by treating the following parameters of the signal as independent:
\begin{eqnarray}
\theta^a&=&\{\alpha,\beta,D_L,t_c,\phi_c,\log M_c, \log \eta\},
\end{eqnarray}
where, as usual, $D_L$ is the luminosity distance to the source, $t_c$, $\phi_c$ are the
time and phase of coalescence, and $M_c$ and $\eta$ are the
chirp mass and symmetric mass ratio of the binary. (For details of the parameter estimation using GR waveforms, see e.g. Ref.~\cite{AISS05}.)
 The errors on $\alpha$ and $\beta$, obtained from the covariance matrix constructed, translate directly into the typical bounds on these parameters using various interferometers.
For the  phasing formula, we use the restricted 3.5PN waveform obtained using PN theory~\cite{BDIWW95,BDEI04} and add the dipole term we propose to the phasing. We ignore the effect of spins, eccentricity and that of harmonics higher than dipole and quadrupole in this work, as a  first approximation.

For the noise spectrum of AdvLIGO detector, we have used the analytic fit
given in Eq.~(2.1) of \cite{MAIS10}. The configuration and the noise model
for third generation Einstein Telescope is based on.
Eq.~(2.2) and (2.3) of \cite{MAIS10}.

The upper frequency cut-off of signals is assumed to be the frequency at the last stable orbit $F_{\rm LSO}$ after which PN approximation ceases to be valid. The quadrupolar term is truncated at $2F_{\rm LSO}$ and the dipole term at $F_{\rm LSO}$, respectively.  The expression for $F_{\rm LSO}$ as a function of the total mass $m$ of the binary, is given by
\begin{equation}
F_{\rm LSO}=(2\pi\,6^{3/2} m)^{-1}.
\end{equation}  
We take the low frequency cut-off of AdvLIGO  to be 20Hz and of ET to be 10Hz. The distance to the source in both cases is fixed to be 200 Mpc. The effect of increasing distance will be to increase the errors in a linear fashion. The mass ratio of the systems is assumed to be 2.

One of the caveats of the present analysis is the use of Fisher matrix as an error estimator. Fisher matrix may underestimate the errors for low signal to noise ratio events\cite{NichVech98,BalDhu98,Vallisneri07}. There have been different
proposals in the literature to go beyond the Fisher matrix~\cite{VZ10,Vallisneri11} in the context of parameter estimation of inspiral signals.
In the present work we do not address these issues and leave them for future.

\vskip 1cm 
\begin{figure}[t]
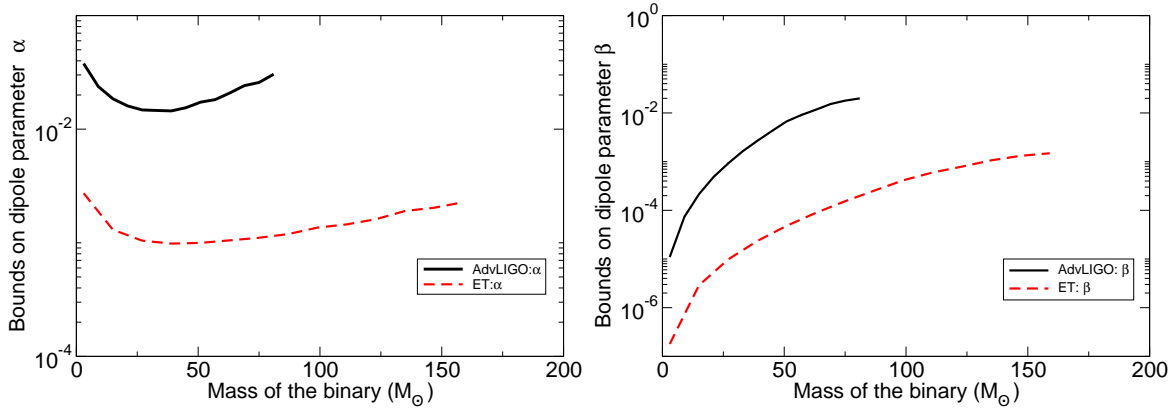

\includegraphics[scale=0.30]{Alpha_AL_ET.eps}
\includegraphics[scale=0.30]{Beta_AL_ET.eps}
\caption{Bounds on $\alpha$ and $\beta$ as a function of the mass of the binary system for a mass ratio of 2.
A lower cut-off in frequency of  10Hz and 20Hz has been used for ET and AdvLIGO, respectively.}\label{VTFisher}
\end{figure}

\section{Results}\label{sec:results}

Fig.~\ref{VTFisher} presents our results. The left panel shows the results for the bounds on $\alpha$ and the right panel for $\beta$ for ground-based interferometers AdvLIGO and ET. The bounds obtained from the Fisher matrix analysis are shown as a function of the total mass of the binary.  The masses range from the binary neutron star or NS-stellar mass black hole (BH) case on the lower side to the black hole binaries consisting of stellar mass BHs or intermediate mass BHs (IMBH) on the higher end.

 The Einstein Telescope would be  able to constrain the dipolar amplitude to $\sim 10^{-3}$ whereas AdvLIGO would bound this amplitude to be $\sim 10^{-2}$. The dipole parameter in the phasing may be best bounded to a value $\sim 10^{-6}$ and $10^{-5}$ by ET and AdvLIGO, respectively. This corresponds to  binary neutron star systems or NS and stellar mass BH systems. However, it should be borne in mind that the event rates for IMBH systems may be far too small at 200 Mpc for AdvLIGO and ET.  Hence the best bounds on $\alpha$ may be weaker by about 5 times if we put the sources at, say 1Gpc, where there may be more sources.  The factor of 5 simply follows from the fact that errors get worse inversely with distance. However for the best bounds on $\beta$, which come from binary NS systems or NS-stellar mass BH systems, 
the estimates based on a luminosity distance of 200 Mpc may not be unrealistic.

From the figures its clear that $\beta$ is always much better estimated 
than the parameter $\alpha$. This can be easily understood because $\beta$
 appears in the phasing, which is more important than the amplitude as far as parameter estimation is concerned. 
  Though there is a $\beta$
dependent contribution in the amplitude also, coming from the re-expansion of ${\dot F}$, it adds much less to the already dominant effect of $\beta$ in the phase.

\section{Conclusion}\label{sec:conclusion}
We proposed a method to constrain the amplitudes of the dipole modes of GW polarization of a generic metric theory of gravity using parameter estimation on detected GW events.
 In the absence of
gravitational waveforms for dipole modes predicted by specific theories of gravity, we wrote down a physically motivated, phenomenological representation of the waveform. Based on  a Fisher-matrix analysis, we conclude that  Advanced ground-based GW detectors like AdvLIGO and ET would provide interesting constraints.

All these analyses have been performed assuming a  single detector configuration. Using a network
of ground-based GW interferometers, one could improve the estimates quoted here. We have not included the effects of higher harmonics, which might again improve the estimates.  It would also be interesting to study the effect of spins and eccentricity on these estimates. We postpone these issues for a future work.

\ack
It is a pleasure to thank Clifford Will, who suggested this problem to me and for comments on the draft. I also thank Nicolas Yunes for
careful reading of an earlier version of this paper and for suggestions and comments. I also thank P Ajith, Bala R Iyer and B S Sathyaprakash for useful discussions and/or comments on the manuscript. I also thank two anonymous referees of this paper whose comments have been valuable in revising the manuscript.
\section{References}
\bibliographystyle{iopart-num.bst}
\bibliography{/home/arun/tphome/arun/ref-list}
\end{document}